\documentclass{IEEEtran}
\IEEEoverridecommandlockouts
\usepackage{amsmath,amssymb,amsfonts}
\usepackage{algorithm}
\usepackage{algpseudocode}
\usepackage{graphicx,subcaption,caption}
\usepackage{textcomp}
\usepackage{xcolor}
\usepackage{mathrsfs}
\usepackage[normalem]{ulem}
\usepackage{cite}

\def\BibTeX{{\rm B\kern-.05em{\sc i\kern-.025em b}\kern-.08em
    T\kern-.1667em\lower.7ex\hbox{E}\kern-.125emX}}

\usepackage{fancyhdr}
\fancypagestyle{firststyle}
{
   \fancyhf{}
   \fancyfoot[C]{Distribution Statement A: Approved for public release. Distribution is unlimited}%

}

\begin{document}

\title{Data-Driven Calibration Technique for \textcolor{black}{Quantitative Radar Imaging}}

\author{\IEEEauthorblockN{Zacharie Idriss and Raghu G. Raj} \\
\IEEEauthorblockA{U.S. Naval Research Laboratory, Washington DC}}
\maketitle
\thispagestyle{firststyle}
\begin{abstract}
   Quantitative inversion algorithms allow for the reconstruction of electrical properties (such as permittivity and conductivity) for every point in a scene. However, \textcolor{black}{such techniques} are challenging to use on measured \textcolor{black}{backscattered phase history signals and datasets} due to the need to know the incident wave field in the scene. In general, this is unknown due to factors such as antenna characteristics, path loss, waveform factors, etc. In this paper, we introduce a scalar calibration factor to account for these factors. To solve for the calibration factor, we augment the inversion procedure by including the forward problem, which we solve by training a simple feed-forward fully connected neural network to learn a mapping between the underlying permittivity distribution and the scattered field at the radar. We then minimize the mismatch between the measured and simulated fields to optimize the scalar calibration factor for each transmitter. \textcolor{black}{We demonstrate the effectiveness of our data-driven calibration approach on the Fresnel Institute dataset wherein we show the accuracy of the estimated scene permittivities. Our paper thus lays the groundwork for the application of quantitative inversion algorithm in real-world imaging scenarios.}
\end{abstract}

\section{Introduction}
In this paper, we present a data-driven calibration technique for quantitative inversion\textcolor{black}{, in particular radar imaging,} algorithms. Specifically, we focus on a contrast-source inversion–based quantitative inversion algorithm called the multiple frequency subspace based optimization method (MFSOM) \textcolor{black}{recently developed by the authors \cite{10576488}.} MFSOM reconstructs the electrical properties (e.g., permittivity) of underlying dielectric scatterers. This can augment scatterer identification algorithms in radar imaging applications, by differentiating materials based on permittivity estimates \cite{10576237,9795338}. Quantitative imaging algorithms such as MFSOM first estimate the induced current density by inverting the scattered wave vector within the domain of interest, and second estimate the permittivity from the current estimate, which requires knowledge of the incident field vector. This makes permittivity estimation challenging as there are two equations; one relating the scattered field to the induced current and one relating the induced current to the permittivity; and three unknowns: the induced current, the permittivity, and the incident field. Overcoming this underdetermined system of data driven calibration is paramount for enabling the application of quantitative imaging algorithms to practical radar imaging scenarios.

The most well-known, automatic, data-driven calibration technique for radar imaging was developed by researchers at the University of Manitoba, which involves solving the forward problem to find a set of weights to apply to the rest of the dataset \cite{6138442}. This technique requires prior knowledge about the scene structure in order to determine the data calibration strategy \cite{Richard_F_Bloemenkamp_2001}. Another set of techniques is rooted in the source reconstruction method, which relies on knowing the total and scattered field at the receiver to find an equivalent current model at the transmitter of the incident field, thus alleviating the need for calibration \cite{768793,5646854,7452542}. Some techniques involve simply knowing the total field \cite{9343673,7735527,electronics8040417,8293231}. Data-driven techniques include introducing scaling factors on the incident wave-field and simultaneously solving for them along with the remaining quantities \cite{YanChang2023,6912945,9632814}. Data-driven techniques using machine learning \cite{9112680,10715281} and compressive sensing \cite{8259463} approaches have also been recently explored in the literature.

In this paper, we augment the original inverse problem by adding a third objective into the underlying cost function; the error between the simulated and scattered wave fields--such that the simulated fields are generated by executing the forward problem. The forward problem involves predicting the electromagnetic fields based on a known model of the medium's properties (e.g., permittivity and conductivity distributions). The simulated wave field is found at each iteration by passing the current contrast function to the forward problem and solving for the scattered wave field. We introduce a calibration factor, $\lambda_c$, that weights both the simulated scattered wave field and the incident field within the state equation. We then use the conjugate gradient descent (CGD) algorithm to find the optimal calibration factor at each iteration. We show that because the incident field is linearly proportional to the scattered field, we can pull out the calibration factor and use it to weight the incident field.

Naturally, solving the forward problem at each iteration is computationally expensive. To overcome this, we train a simple feed-forward, fully-connected, neural network (FCNN) to learn a forward operator between the scene and the scattered wave field. Using this pre-trained network, we introduce the forward problem in our inverse algorithm without any significant computational cost. We observe that by training the FCNN using a variety of examples, we are able to overcome the need to solve the forward problem in an exact manner. \textcolor{black}{We demonstrate the effectiveness of our data-driven quantitative imaging algorithm using the Fresnel Institute dataset \cite{Geffrin_2005}.} 

In our equations, uppercase symbols denote operators or large-scale functions (e.g. \( \mathcal{M} \)), while lowercase symbols represent physical fields or parameters at a specific point (e.g., \( \mathbf{r} \) for position). \textcolor{black}{Boldface symbols (e.g., \( \mathbf{E} \) and \( \mathbf{W} \)) indicate matrix representations of vector fields}, while non-bold symbols represent \textcolor{black}{either scalar quantities (e.g., \( \omega \) for angular frequency and \( \chi \) for a scalar field) or vector fields (e.g. $\mathscr{E}$ and $\mathcal{W}$).}

\section{Forward Problem}
Let $S\subset\mathbb{R}^2$ denote the open and bounded sensor domain. We then require that all scatterers are localized in an open and bounded domain \( D \subset \mathbb{R}^2 \) satisfying \(D \cap S = \emptyset\). We start with the Lippmann–Schwinger equation for the total electric field  \cite{chen2018computational} \( \mathscr{E}(\mathbf{r}):D\rightarrow\mathbb{C}\) in volts per meter \(\left[\text{V}/\text{m}\right]\),
\begin{equation}\label{eq:lippschw}
\mathscr{E}(\mathbf{r}) = \mathscr{E}_{\text{inc}}(\mathbf{r}) + j\omega\mu\int_D \mathbf{g}(\mathbf{r}, \mathbf{r}') \mathcal{W}(\mathbf{r}') \, d\mathbf{r}',
\end{equation}
where \( \mathscr{E}_{\text{inc}}(\mathbf{r}):S\times D\rightarrow\mathbb{C}\) is the incident electric field in volts per meter \(\left[\text{V}/\text{m}\right]\), \(D\) is the integration domain in square meters \(\left[\text{m}^2\right]\), let \(G(\mathbf{r},\mathbf{r}')\colon D\times D\to\mathbb{C}\) denote the Green’s function, with units of inverse length \(\left[\text{m}^{-1}\right]\), and \( \mathcal{W}(\mathbf{r}'):S\times D\rightarrow\mathbb{C}\) is the current density in amperes per square meter \(\left[\text{A}/\text{m}^2\right]\). In general, \eqref{eq:lippschw} is expressed in matrix form as:
\begin{equation}\label{eq:lippschw2}
\mathbf{E} = \mathbf{E}_{\text{inc}} + \mathbf{G}\mathbf{W},
\end{equation}
where, $\mathbf{E}$, $\mathbf{E}_{\text{inc}}$, $\mathbf{G}$, and $\mathbf{W}$ are the matrix representations of $\mathscr{E}$, $\mathscr{E}_{\text{inc}}$, $\mathbf{g}$, and $\mathcal{W}$ respectively, which are determined through the Method of Moments using a pulse basis function expansion and point collocation (Dirac delta distributions) \cite{harrington1968}. \textcolor{black}{From \eqref{eq:lippschw} and using the constitutive relations for linear isotropic media \cite{balanis2012advanced}, we define an invertible linear operator \( \mathcal{M} : \mathcal{W}\rightarrow\mathscr{E}_{\text{inc}}\) as}
\begin{equation}\label{eq:op1}
    \mathcal{M}[\mathcal{W}](\mathbf{r}) = \frac{\mathcal{W}(\mathbf{r})}{i \omega \epsilon_0 \chi(\mathbf{r})} - j\omega\mu\int_V \mathbf{g}(\mathbf{r}, \mathbf{r}') \mathcal{W}(\mathbf{r}') \, d\mathbf{r}'
\end{equation}
where \( \omega \) is the angular frequency in radians per second (rad/s), \( \epsilon_0 \) is the permittivity of free space in farads per meter (F/m), and \( \chi(\mathbf{r}) :D\rightarrow\mathbb{C}\) is the susceptibility, a dimensionless parameter. 

Although the operator $\mathcal{M}$ is invertible, computing its inverse explicitly is prohibitively expensive for any reasonably sized domain,~$D$. In this paper, we therefore use CGD to avoid forming~$\mathcal{M}^{-1}$ directly, which significantly accelerates the computation.

Expressing \eqref{eq:op1} with $\lambda_c \in \mathbb{C}$, the unknown calibration factor to be determined, we have,
\begin{equation}\label{eq:fwd}
    \mathcal{M}[\mathcal{W}](\mathbf{r}) = \lambda_c \mathscr{E}_{\text{inc}}(\mathbf{r}).
\end{equation}

Solving for \( \mathcal{W}(\mathbf{r}) \) explicitly, we obtain
\[
\mathcal{W}(\mathbf{r}) = \lambda_c \mathcal{M}^{-1} \mathscr{E}_{\text{inc}}(\mathbf{r}),
\]
showing the current density \( \mathcal{W}(\mathbf{r}) \) in terms of the inverse of the operator \( \mathcal{M} \) applied to the scaled incident field \( \lambda_c \mathscr{E}_{\text{inc}}(\mathbf{r}) \). Thus, the scattered field is expressed, using Green's Theorem, as
\begin{equation}\label{eq:scatE}
    \mathscr{E}_{\text{scat}}(\mathbf{r}_{\text{rec}}) = \lambda_c \int_V \mathbf{g}(\mathbf{r}_{\text{rec}}, \mathbf{r}') \, \mathcal{M}^{-1} \mathscr{E}_{\text{inc}}(\mathbf{r}') \, d\mathbf{r}'.
\end{equation}
Equation \eqref{eq:scatE} is expressed in matrix form as:
\begin{equation}\label{eq:scatE2}
\mathbf{E}_s = \lambda_c\mathbf{G}\mathbf{W},
\end{equation}
where $\mathbf{E}_s$ is the matrix representation of $\mathscr{E}_{\text{scat}}$.

\textcolor{black}{\section{Data-driven Calibration}}

In the \textcolor{black}{MFSOM algorithm \cite{10576488}}   , the current is separated into distinct subspaces—a signal and a noise subspace—using the singular value decomposition (SVD) of the Green’s functions as
\[
\mathbf{G} = \mathbf{U}\mathbf{\Sigma}\mathbf{V}^*
\]
where the matrices $\mathbf{U}$, $\mathbf{\Sigma}$, and $\mathbf{V}$ denote the left singular vectors, singular values, and right singular vectors, respectively. Using this formulation, the cost function is presented as
\begin{align}\label{eq:csi}
    J(\mathbf{W}, \chi) &= \sum_{k=1}^{K}\sum_{p=1}^{P} \frac{1}{2}\Bigg( \frac{\| \mathbf{E}_{s}^{k,p} - \mathbf{G} \mathbf{W}^{k,p} \|_2^2}{\| \mathbf{E}_s^{k,p} \|_2^2}\nonumber\\
    &+\frac{\| \mathbf{W}^{k,p} - \chi \left(\mathbf{E}_\text{inc}^{k,p} + \mathbf{E}_d^{k,p}\right) \|^2_2}{\| \mathbf{W}^{k,p}_+ \|_2^2} \Bigg)
\end{align}
where \( \mathbf{E}_s^{k,p}\) denotes the measured scattered field for the \(p\)-th incident field at frequency \(k\), \(\mathbf{E}_d^{k,p} \) denotes the scattered field within the domain of interest (i.e., when both \(\mathbf{r},\mathbf{r}'\) in \eqref{eq:scatE} denote points in the scene), \( \mathbf{G} \) is Green's function, the dominant current subspace is defined as
\[
\mathbf{W}^{k,p}_+ = \mathbf{V}^{k,p}_+ \boldsymbol{w}^{k,p}_+,\quad\boldsymbol{w}^{k,p}_+=\frac{\mathbf{U}^*_+\mathbf{E}_s^{k,p}}{\mathbf{\Sigma}_+}
\]
for the \( p \)-th incident field at frequency \(k\), the subscript $+$ denotes the signal subspace, and \( \mathbf{E}_\text{inc}^{k,p} \) is the incident field for the \( p \)-th source at frequency \(k\). \textcolor{black}{We further remark that the first term in equation \eqref{eq:csi} is called the data-fidelity term, whereas the second term in \eqref{eq:csi} is called the state equation mismatch term}.

As seen from \eqref{eq:csi}, the incident field in the state equation mismatch is scaled by the scalar field, $\chi$. However, in \textcolor{black}{received backscattered signals or measured datasets}, the incident field is scaled by unknown factors such as antenna beam pattern, antenna gain, and waveform amplitudes, which have to be estimated to accurately reconstruct $\chi$. To account for all these effects, we optimize for the scalar calibration factor \textcolor{black}{(for each calibration angle)}, $\lambda_c^p$, and augment the problem \eqref{eq:csi} as
\begin{align}\label{eq:csi_2}
    J(\mathbf{W}, \chi, \{\lambda_c\}_{k,p=1}^{K,P})& = \sum_{k=1}^{K}\sum_{p=1}^{P} \frac{1}{2}\Bigg(
    \frac{\| \mathbf{E}_s^{{\text{meas}},k,p} - \mathbf{G} \mathbf{W}^{k,p} \|_2^2}{\| \mathbf{E}_s^{{\text{meas}},k,p} \|_2^2} \nonumber \\
    &+ \frac{\| \mathbf{W}^{k,p} - \chi\left(\lambda_c^{k,p}  \mathbf{E}_\text{inc}^{k,p}+ \mathbf{E}_d{k,p} \right)\|_2^2}{\| \mathbf{W}^{k,p}_+ \|_2^2} \nonumber \\
    &+  \textcolor{blue}{\frac{\left\| \lambda_c^{k,p} \mathbf{E}_s^{\text{sim},k,p} - \mathbf{E}_s^{{\text{meas}},k,p} \right\|_2^2}{\left\| \mathbf{E}_s^{{\text{meas}},k,p} \right\|_2^2} + \beta |\lambda_c^{k,p}|^2}\Bigg)
\end{align}
where \( \mathbf{E}_s^{\text{sim}} \) represents the simulated field vector, \( \mathbf{E}_s^{\text{meas}} \) is the measured field vector, and \( \beta \in \mathbb{R} \) is the regularization parameter balancing data fidelity and smoothness. As shown in \eqref{eq:scatE}, the added calibration factor that scales the incident field in the forward operator is factored out as $\mathcal{M}$ is a linear operator. Thus, minimizing the third term in \eqref{eq:csi_2} with respect to $\lambda_c$ provides the correct scaling, used to scale the incident field in the second term in \eqref{eq:csi_2}. Note that in the following sections, for notational clarity, we drop the dependence on \(k\); however, the analysis still holds, and the procedures defined by the above cost functions are used in the numerical results.

In \eqref{eq:csi_2}, \( \mathbf{E}_s^{\text{sim}} \) is found by first solving \eqref{eq:fwd}, which can be computationally expensive for a large domain, high dielectric constant, or large number of frequencies. Therefore, as detailed in Subsection \ref{subsec:FCNN}, we train an FCNN to learn a mapping (shown in Fig. \ref{fig:nnfwd}) that enables us to solve for the scattered field much faster once the FCNN is trained. \textcolor{black}{Importantly, the central benefit of this approach is that the resulting algorithm can be applied to a received backscattered signal without the need for knowing the scene structure in advance.}

To find the optimal \( \lambda_c \in \mathbb{C} \) that minimizes the error between the measured and simulated fields, we define the objective as
\[
\{\lambda_c\}_{p=1}^P = \arg \min_{\{\lambda_c\}_{p=1}^P\in\mathbb{C}} J(\mathbf{W}, \chi, \{\lambda_c\}_{p=1}^P).
\]
CGD is used to solve for \( \lambda_c \), where search directions are found using the gradient \( g = \nabla_{\bar{\lambda}_c} J \) to inform an \textit{adaptive} search direction, $d$, using the \textcolor{black}{Polak–Ribiere update \cite{nocedal2006numerical}.} Then, $\lambda_c$ is found in an iterative manner using the following update rule
\[
\lambda_c^{p,(k+1)} = \lambda_c^{p,(k)} + \alpha^{p,(k)} d^{p,(k)}
\]
where \( \alpha^{p,(k)} \in\mathbb{C}\) is the step size that minimizes the cost function along \( d^{p,(k)}  \in\mathbb{C}\), and is found using line search minimization. Thus, at each iteration, $k$, the sub objective of finding the step size, \( \alpha \), is to minimize the cost function 
\[
\alpha^{p,(k)} = \arg \min_{\alpha\in\mathbb{C}} J(\mathbf{W}, \chi, \lambda_c^{p,(k)} + \alpha d^{p,(k)})
\]
along, \( d^{p,(k)} \). This objective is minimized by differentiating with respect to \( \alpha \) and solving for the optimal \( \alpha^{p,(k)} \) in closed form. \textcolor{black}{The pseudo-code of the resulting Quantitative Radar Imaging (QRI) algorithm that incorporates the above steps is presented in Algorithm \ref{alg:1}. Thereafter, sub-section \ref{subsec:FCNN} details the construction of the neural network used in Algorithm \ref{alg:1} for automatic, data-driven calibration and section \ref{sec:numres} details the numerical results of the QRI algorithm. In Algorithm \ref{alg:1}, $T$ denotes the termination tolerance, and for every result in this paper, it was set to $T=5\times10^{-4}$.}

\begin{algorithm}
    \caption{\textcolor{black}{QRI with Data-driven Calibration}}
    \begin{algorithmic}
        \State \textbf{Input:} Define $\mathbf{W}_0$, $\chi_0,\beta$, $\{\lambda_{c,0}\}_{p=1}^P=1,n=1$
        \State \textbf{Initialize:} $J_{\text{prev}} \gets \infty$ \Comment{Set initial cost function value}
        \While{$|J(\mathbf{W}, \chi, \lambda_c) - J_{\text{prev}}| > T$} \Comment{Termination criterion}
            \State $J_{\text{prev}} \gets J(\mathbf{W}, \chi, \{\lambda_c\}_{p=1}^P)$
            \vspace{1em}
            \State $\mathbf{W}_n \gets \arg \min_{\mathbf{W}} J(\mathbf{W}, \chi, \{\lambda_c\}_{p=1}^P)$ \Comment{Induced current}
            \State Solve using CGD
            \vspace{1em}
            \State $\chi_n \gets \arg \min_{\chi} J(\mathbf{W}, \chi,\{\lambda_c\}_{p=1}^P)$ \Comment{Contrast function}
            \State Solve using CGD
            \vspace{1em}
            \State \textbf{Update} $\mathbf{E}_s^{\text{sim}}(\chi_n)$ \Comment{Update simulated field}
            \State Using exact forward problem or trained FCNN
            \vspace{1em}
            \State $\{\lambda_{c,n}\}_{p=1}^P \gets \arg \min_{\lambda_c} J(\mathbf{W}, \chi, \{\lambda_c\}_{p=1}^P)$ \Comment{Scaling parameter}
            \State Solve using CGD
            \State $n=n+1$ \Comment{Update iteration count}
        \EndWhile
        \State \textbf{Output:} Optimized values of $\mathbf{W}$, $\chi$, $\{\lambda_c\}_{p=1}^P$
    \end{algorithmic}
    \label{alg:1}
\end{algorithm}

\subsection{Approximate Forward Problem using Neural Network}\label{subsec:FCNN}
\begin{figure}
    \centering
    \includegraphics[width=0.9\linewidth]{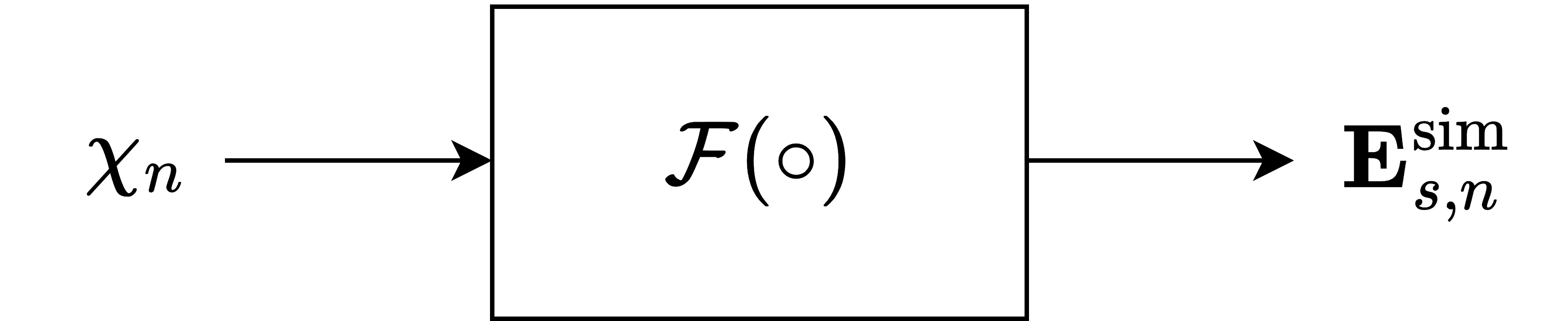}
    \caption{Flow map for input-output relationship used to train the FCNN. Input: estimated $\chi$. Output: simulated wave field $\mathbb{E}^{\text{sim}}_s$. The subscript $n$ is used to show that this occurs at every iteration of the algorithm.}
    \label{fig:nnfwd}
\end{figure}

We train a neural network to learn an arbitrary mapping, $\mathcal{F}$\textcolor{black}{, as shown in Fig. \ref{fig:nnfwd}, }
from the estimated permittivity distribution, $\epsilon_R(\mathbf{r})\in\mathbb{R}$, to the scattered wave field at the receiver, $\mathbf{E}_{s}^{\text{sim}}(\mathbf{r})\in\mathbb{C}$. The mapping from the estimated \( \chi \) \( \mathbf{E}_s^{\text{sim}} \) is expressed as
\[
\mathcal{F}: \chi \to \mathbf{E}_s^{\text{sim}},
\]
where \( \mathcal{F} \) denotes the forward operator that generates \( \mathbf{E}_s^{\text{sim}} \) from a given \( \chi \).

To \textcolor{black}{construct an optimal} neural network with training data, the forward problem was run 100 times for each scattering configuration shown in Fig. \ref{fig:configs}. Diversity in the training samples was introduced by placing a uniform distribution over the permittivity and assigning each homogeneous scatterer a value chosen as $\epsilon_R \sim \text{Uniform}(1.1, 5)$, where Uniform denotes the uniform distribution. A lower bound was placed on $\epsilon_R$ at 1.1 to ensure that each scatterer was sufficiently differentiated from the background and not smaller than \textcolor{black}{unity (which corresponds to the relative permittivity of free space)}. 

\begin{figure}
    \centering
    \includegraphics[width=0.9\linewidth]{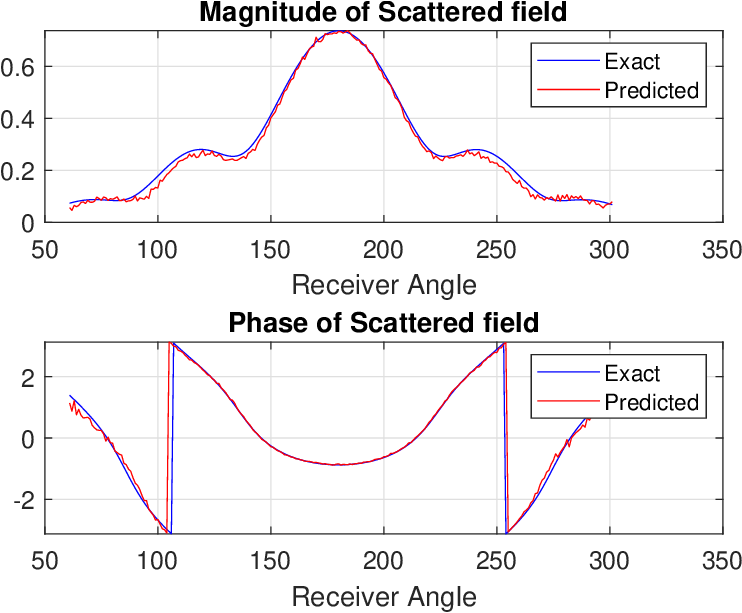}
    \caption{Predicted versus measured wave field $\mathbf{E}_s^{\text{meas}}$, for \textcolor{black}{the scene in Fig. \ref{fig:FoamDielExt}} using the pretrained FCNN. Single transmitter at 0 degrees.}
    \label{fig:NN_predict}
\end{figure}

In Fig. \ref{fig:NN_predict} we show that using the pre-trained FCNN for the exact scene shown in Fig. \ref{fig:FoamDielExt}, we are able to achieve agreement between the exact and predicted wave fields with a mean square error of $\sim8.4\times 10^{-4}$.
\section{Numerical Results}\label{sec:numres}
\begin{figure}
    \centering   
    \begin{subfigure}[b]{0.45\textwidth}
        \centering
        \includegraphics[width=0.6\columnwidth]{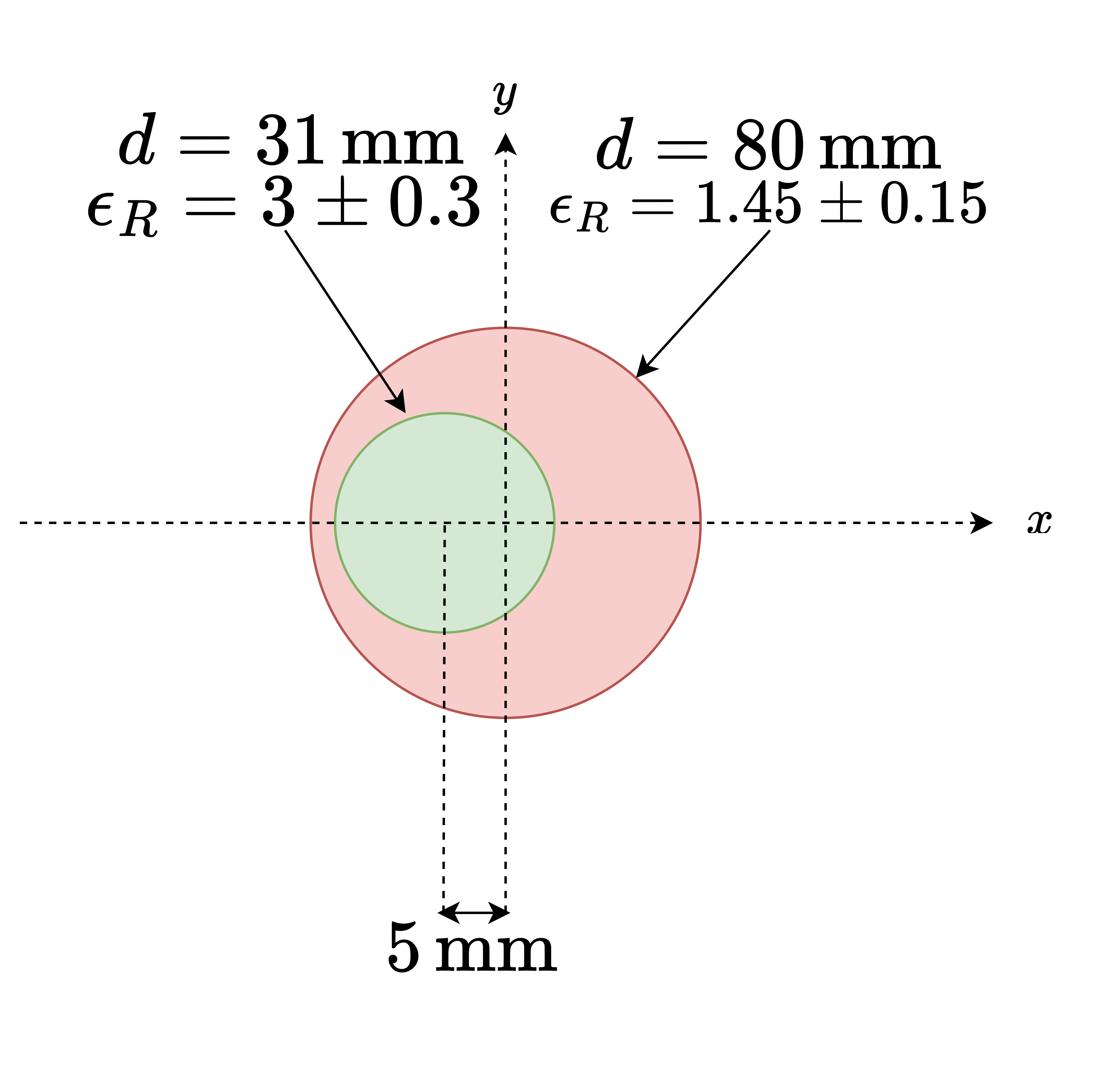}
        \caption{`FoamDielInt': a foam cylinder (red) with a plastic cylinder (berylon) (green) within it.}
        \label{fig:FoamDielInt}
    \end{subfigure}
    \hfill
    \begin{subfigure}[b]{0.45\textwidth}
        \centering
        \includegraphics[width=0.6\columnwidth]{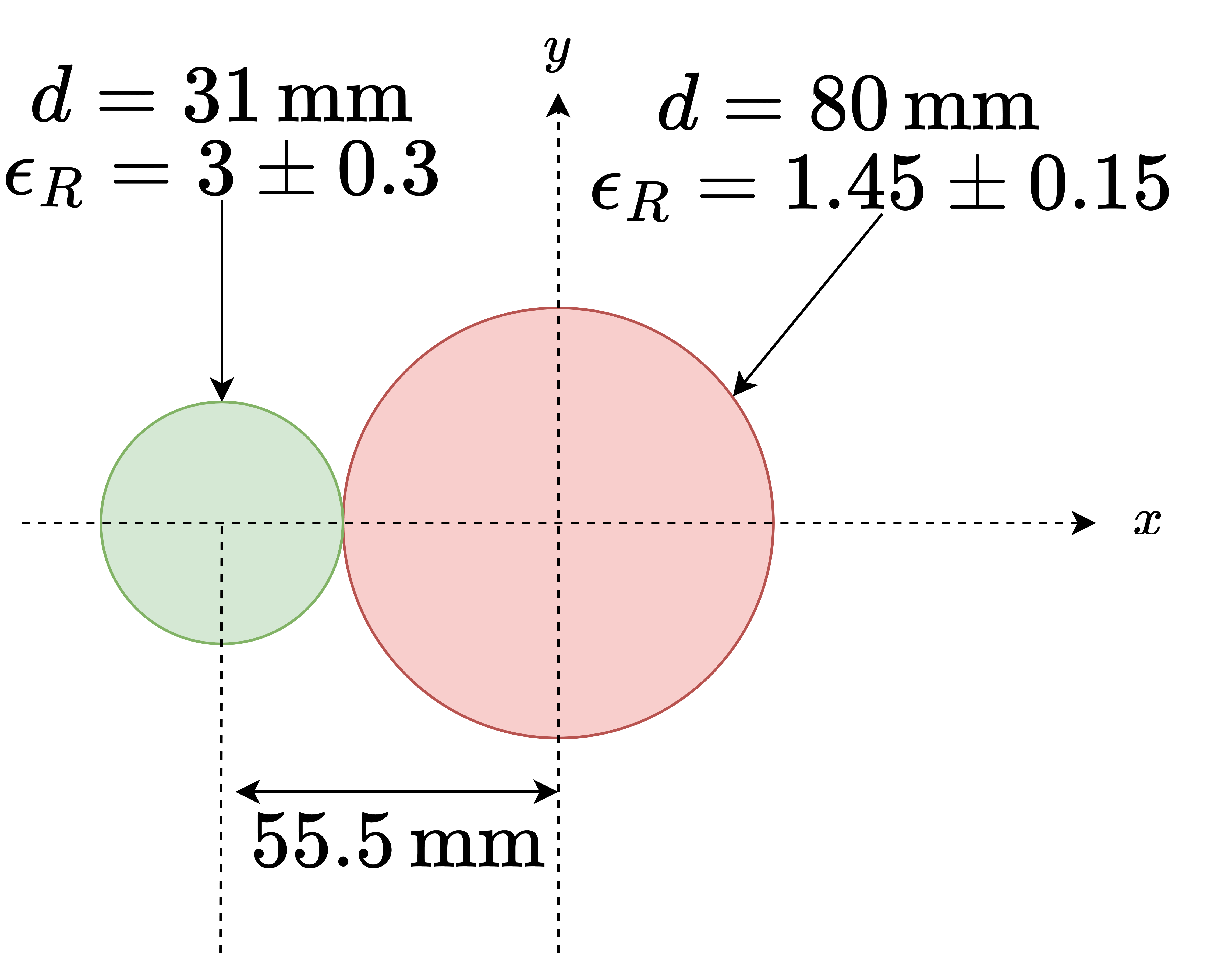}
        \caption{`FoamDielExt': a foam cylinder (red) with an external plastic cylinder (berylon) (green).}
        \label{fig:FoamDielExt}
    \end{subfigure}
    \caption{Experimental scattering configurations from the Institute Fresnel dataset \cite{Geffrin_2005}. In Figs. \ref{fig:FoamDielInt} and \ref{fig:FoamDielExt}, $d$ denotes diameter, and $\epsilon_R$ denotes relative permittivity.}
    \label{fig:configs}
\end{figure}

We test our calibration algorithm on the configurations shown in Fig. \ref{fig:configs}. The configuration in Fig. \ref{fig:FoamDielInt} consists of two dielectric scatterers: the red foam with a relative permittivity of $\epsilon_R=1.45\pm0.15$ and the green plastic with a relative permittivity of $\epsilon_R=3\pm0.3$. The configuration in Fig. \ref{fig:FoamDielExt} features the plastic tube now adjacent to the red foam scatterer.

We place a metric on the reconstruction performance of our calibration algorithm using the normalized squared error (NSE) defined as
\begin{equation}\label{eq:NSE}
    \Delta = \sum_{l=1}^L \frac{\lvert \hat{\epsilon}_l - \epsilon_l \rvert^2}{\lvert \epsilon_l \rvert^2}
\end{equation}
where $\hat{\epsilon}_l$ denotes the estimated permittivity at pixel $l$, $\epsilon_l$ represents the ground truth value at pixel $l$, and $L$ is the total number of image pixels. 

\begin{figure}
    \centering
    \includegraphics[width=1\linewidth]{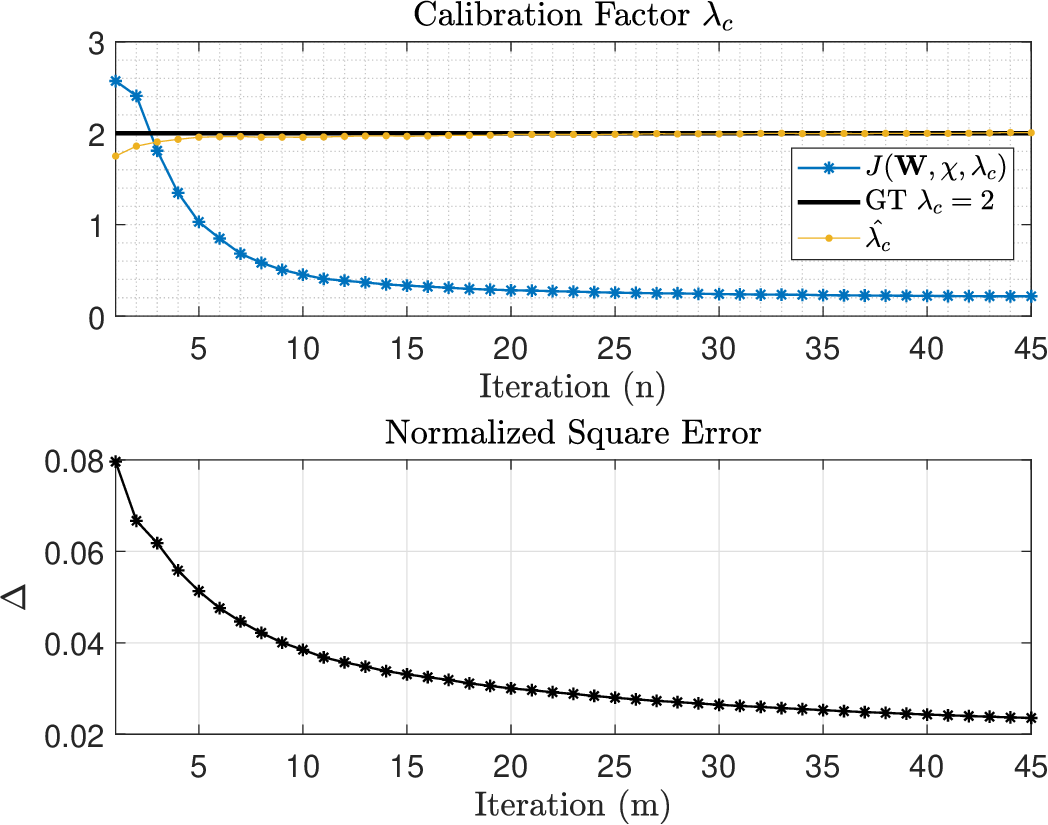}
    \caption{Jointly estimated calibration factor for all transmitters $p, p=1,\dots,8$. \textit{Top}: The cost function value is shown in blue, and the GT value of $\lambda_c=2$ is shown in black. The yellow line corresponds to the joint estimation of $\lambda_c$ for all transmitters. The regularization parameter is set to $\beta=1\times 10^{-3}$. \textit{Bottom}: NSE where $\Delta$ denotes the normalized squared error and is defined in \eqref{eq:NSE}.}
    \label{fig:w_2_beta_1e-3}
\end{figure}

We first assign a ground truth (GT) calibration factor as $\lambda_c^p=\lambda_c=2$ i.e., each transmitter is the same, and consider the configuration shown in Fig. \ref{fig:FoamDielExt}. Here, $p=1,\dots,8$. We then use the \textcolor{black}{procedure} shown in Algorithm \ref{alg:1} to estimate the calibration factor from $\mathbf{E}^{\text{meas}}_s$, at each receiver, \textcolor{black}{together with the  induced current, and estimated $\chi$.} To present the results of our algorithm, we show the estimated scalar calibration factor in Fig. \ref{fig:w_2_beta_1e-3} as a function of iterations, $n$, to demonstrate convergence of our estimate to the GT value along with the overall cost function shown computed as \eqref{eq:csi_2}.

\textcolor{black}{Since $\lambda_c^p = \lambda_c = 2$, a single calibration factor is jointly estimated as a function of all $p$ by coherently summing the gradient and step size across $p$. This approach yields good results when the underlying assumption holds. However, as discussed in Subsection~\ref{subsuc:exp}, this assumption is invalid when processing experimental datasets due to unknown channel effects at each angle, which are absent in simulations.} 

The estimated calibration factor, denoted as $\hat{\lambda_c}$, is shown in Fig. \ref{fig:w_2_beta_1e-3} using a yellow line with circle markers. The convergence of the cost function, $J(\mathbf{W},\chi,\lambda_c)$, is shown in blue with asterisk markers. The GT value for $\lambda_c$ is shown in black. We show that the data-driven estimation of the calibration factor approaches the GT value. To estimate $\lambda_c$ in Fig. \ref{fig:w_2_beta_1e-3}, we use $\beta=1\times 10^{-3}$. However, the results depend on a good choice for $\beta$, which regularizes the estimation of $\lambda_c$ as shown in \eqref{eq:csi_2}, and Algorithm \ref{alg:1}. Determining the optimal value of $\beta$ automatically, is not straightforward, and the choice significantly affects the fidelity of the results. 

\begin{figure}
    \centering
    \includegraphics[width=1\linewidth]{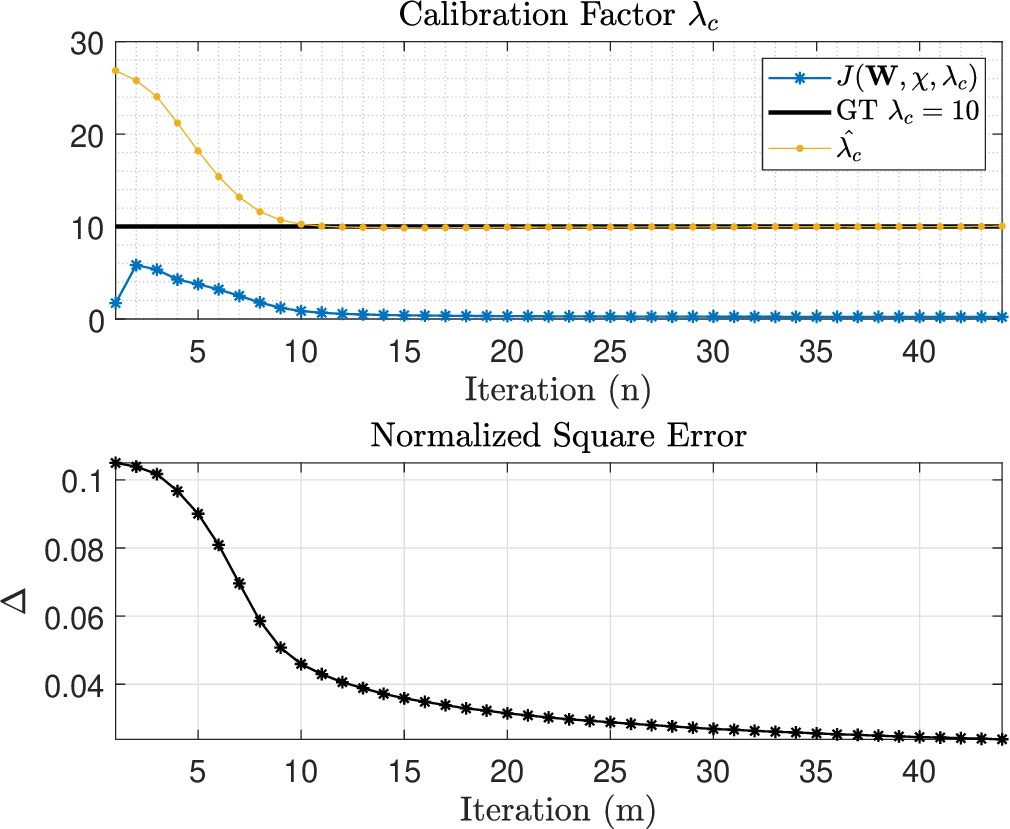}
    \caption{Jointly estimated calibration factor for all transmitters $p, p=1,\dots,8$. \textit{Top}: The cost function value is shown in blue, and the GT value of $\lambda_c=10$ is shown in black. The yellow line corresponds to the joint estimation of $\lambda_c$ for all transmitters. The regularization parameter is set to $\beta=1\times 10^{-5}$. \textit{Bottom}: NSE where $\Delta$ denotes the normalized squared error and is defined in \eqref{eq:NSE}.}
    \label{fig:w_10_beta_1e-5}
\end{figure}

We now consider the case for a GT value of $\lambda_c^p=\lambda_c=10$ i.e., each transmitter is the same. The results are shown in Fig. \ref{fig:w_10_beta_1e-5} with $\beta=1\times 10^{-5}$, where the \textit{top} figure shows the convergence of the cost function in blue, the GT $\lambda_c$ is shown in black, and $\hat{\lambda_c}$, is shown in yellow. 

\subsection{Experimental Dataset}\label{subsuc:exp}
\begin{figure}
    \centering   
    \begin{subfigure}[b]{0.45\textwidth}
        \centering
        \includegraphics[width=1\columnwidth]{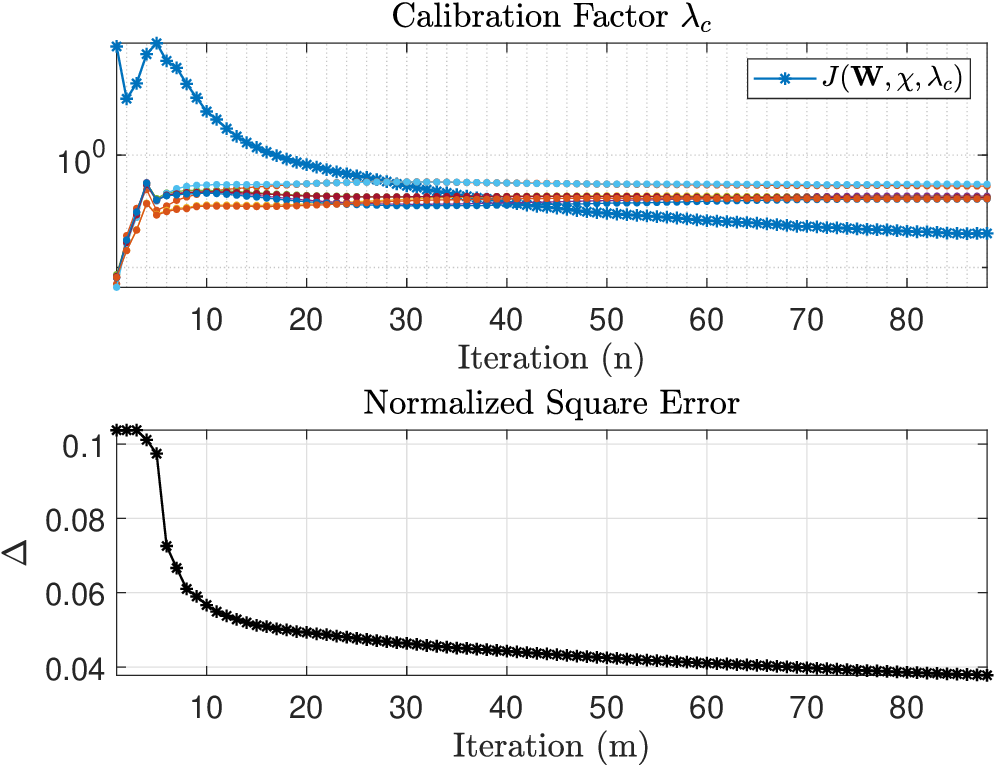}
        \caption{\textit{Top}: The estimated calibration factors, $\lambda_c$, for the measured dataset (Fig.~\ref{fig:FoamDielExt}) from \cite{Geffrin_2005} for Tx$_{1}$ to Tx$_{8}$, are presented on a logarithmic scale. The overall cost function is shown in blue. Here, $\beta=1$. \textit{Bottom}: NSE where $\Delta$ denotes the normalized squared error and is defined in \eqref{eq:NSE}.}
        \label{fig:DielExtTM_calib}
    \end{subfigure}
    \hfill
    \begin{subfigure}[b]{0.45\textwidth}
        \centering
        \includegraphics[width=1\columnwidth]{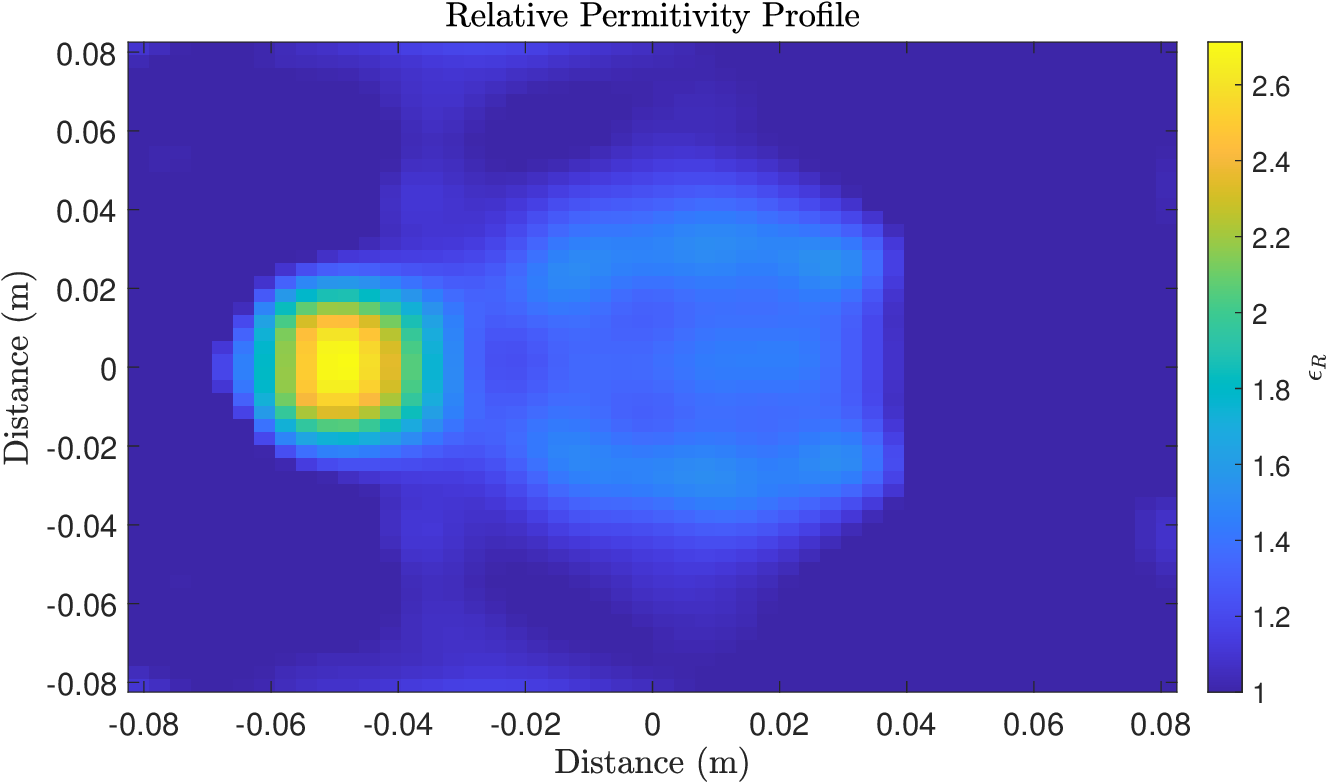}
        \caption{The corresponding permittivity distribution for the calibration in Fig. \ref{fig:DielExtTM_calib}. The NSE is 0.038.}
        \label{fig:FoamDielExt_calib_perm}
    \end{subfigure}
    \caption{Estimated calibration factors and reconstructed permittivity distribution for the scattering configuration in Fig. \ref{fig:FoamDielExt}.}
    \label{fig:calib_results}
\end{figure}

We now test our calibration algorithm on the dataset obtained from the Fresnel Institute \cite{Geffrin_2005}. We consider the scattered wave field data corresponding to the scattering configuration shown in Fig.~\ref{fig:FoamDielExt}. The dataset consists of $P = 8$ transmitters (Tx) and $Q = 241$ receivers, all located at an equal distance of 1.67 meters from the scene center, located at (0,0) in Fig. \ref{fig:FoamDielExt_calib_perm}.

Fig. \ref{fig:calib_results} shows the results of applying our calibration to the measured data corresponding to Fig. \ref{fig:FoamDielExt}. 
Fig.~\ref{fig:DielExtTM_calib} shows the absolute values of the calibration factors for Tx$_{1}$ to Tx$_{8}$ in colored lines with circle markers (note that some of the colored lines obscure each other and are not all visible), and the value of the cost function in \eqref{eq:csi_2} for each iteration, \(n\), in blue with asterisk markers. We allow the scalar calibration factor for each Tx to be complex to account for phase factors and find that a purely real calibration factor does not yield a good NSE. Moreover, as can be seen from Fig. \ref{fig:DielExtTM_calib}, we do not assume $\lambda_c^p=\lambda_c$, and when we test this, the joint estimation yields inferior results. Therefore, it is noted that the assumption that each transmitter is the same does not always hold and that the more general form of the algorithm should be used when such a prior assumption cannot be made. Correspondingly, we present the resulting scene in Fig.~\ref{fig:FoamDielExt_calib_perm}, where the color bar on the right-hand side indicates the permittivity value for each pixel. Table~\ref{tab:001} displays the NSE for the reconstruction performance and compares our calibration Algorithm~\ref{alg:1} to the NSE when not calibrating the dataset. In Table \ref{tab:001}, the title \textit{Yes} denotes that the calibration Algorithm \ref{alg:1} was used, and the title \textit{No} denotes that $\lambda_c=1\forall p$, i.e. the calibration factor was forced to 1 at each iteration. This result demonstrates the effectiveness and potential of the QRI algorithm for performing accurate quantitative radar imaging on measured datasets in a data-driven way.
\begin{table}
    \centering
    \caption{\textsc{Reconstruction Error}}
    \begin{tabular}{|c|c|c|}
    \hline
     Calibrated & Yes & No \\ \hline
     NSE $(\Delta)$ & 0.038 & 0.201\\ \hline
    \end{tabular}    
    \label{tab:001}
\end{table}

\begin{figure}
    \centering
    \includegraphics[width=1\linewidth]{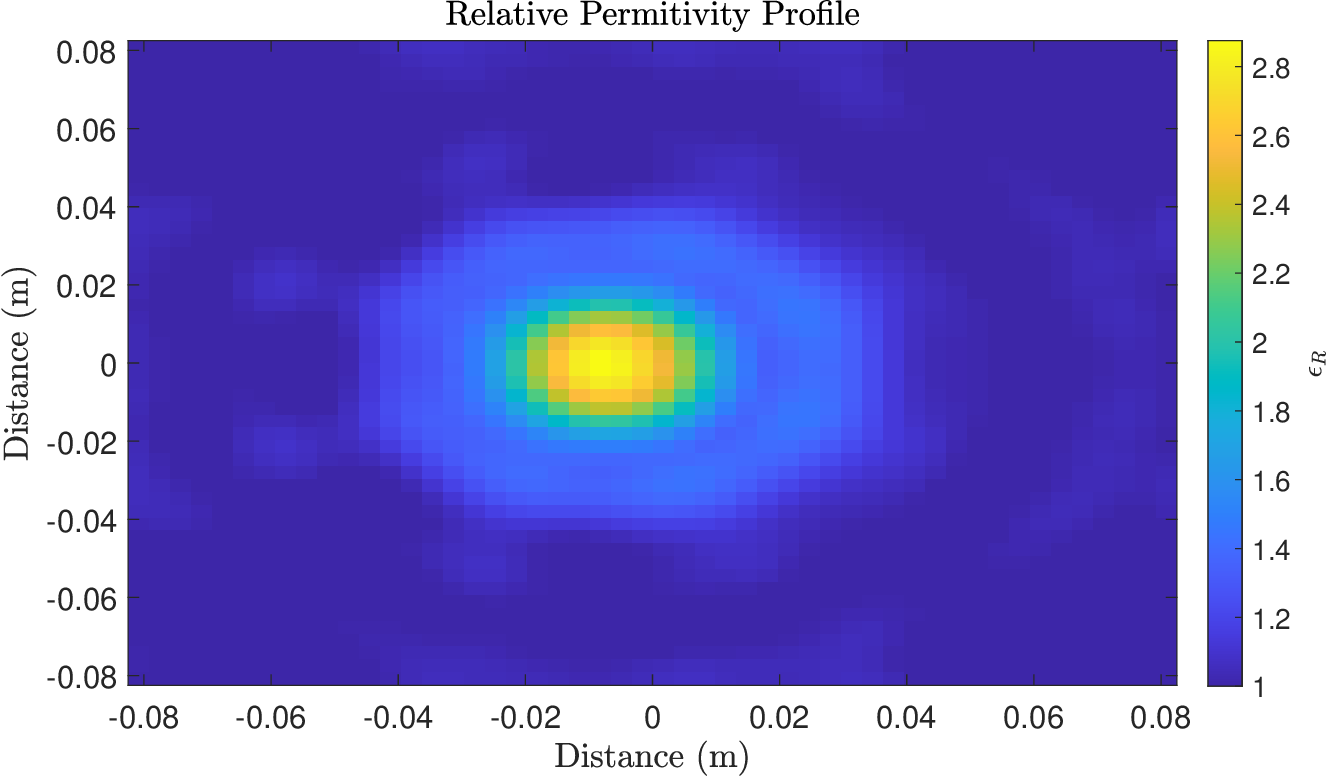}
    \caption{The estimated relative permittivity profile using the MFSOM algorithm with data-driven calibration, for the scattering configuration shown in Fig. \ref{fig:FoamDielInt}.}
    \label{fig:extra}
\end{figure}

We further present the results for the configuration shown in Fig. \ref{fig:FoamDielInt} in Fig. \ref{fig:extra}.

\section{Conclusion}
We find that our data-driven calibration technique is able to estimate a scalar calibration factor for each transmitter by incorporating the forward problem into the inverse problem. Since this is computationally expensive, we use an FCNN to avoid having to solve the forward problem exactly at each iteration. \textcolor{black}{We note that adding the forward problem can provide a robust solution to estimating the incident field. An important future challenge arising from this paper is the need to alleviate the dependence of our current QRI algorithm on choosing the right $\beta$. This can be realized by automatically adapting $\beta$ within a neural network processing framework. The resulting algorithm should find wide-ranging applications in quantitative radar imaging including, for example, applications to ground-penetrating radar imaging for estimating the permittivity of the ground and scatterers.}

\bibliographystyle{IEEEtran}
\bibliography{report2}

\end{document}